\documentclass[preprint2]{aastex}
\usepackage{graphicx}

\shorttitle{CHARGE-EXCHANGE-INDUCED X-RAY EMISSION}
\shortauthors{Ali et al.}

\begin{document}


\title{CRITICAL TEST OF SIMULATIONS OF CHARGE-EXCHANGE-INDUCED X-RAY EMISSION IN THE SOLAR SYSTEM}


\author{R. Ali,\altaffilmark{1} P. A. Neill,\altaffilmark{2} P.
Beiersdorfer,\altaffilmark{3} C. L. Harris,\altaffilmark{4} D. R. Schultz,\altaffilmark{5} and P. C. 
Stancil\altaffilmark{6}}


\altaffiltext{1}{Department of Physics, The University of Jordan, Amman 11942, Jordan; 
ramimali@ju.edu.jo}
\altaffiltext{2}{Department of Physics, University of Nevada, Reno, NV 89557,
USA; paul@physics.unr.edu}
\altaffiltext{3}{Lawrence Livermore National Laboratory, 7000 East Avenue, L-260,
Livermore, CA 94550, USA; beiersdorfer@llnl.gov}
\altaffiltext{4}{Natural Sciences Division, Gulf Coast Community College, Panama
City, FL 32401, USA; charris@ gulfcoast.edu}
\altaffiltext{5}{Physics Division, Oak Ridge National Laboratory, Oak Ridge, TN 37831,
USA; schultzd@ornl.gov}
\altaffiltext{6}{Department of Physics and Astronomy, University of Georgia, Athens,
GA 30602, USA; stancil@physast.uga.edu}


\begin{abstract}
Experimental and theoretical state-selective X-ray spectra resulting from single-electron capture in charge exchange (CX) collisions of Ne$^{10+}$ with He, Ne, and Ar are presented for a collision velocity of 933 km s$^{-1}$ (4.54 keV nucleon$^{-1}$), comparable to the highest velocity components of the fast solar wind.  The experimental spectra were obtained by detecting scattered  projectiles, target recoil ions, and X-rays in coincidence; with simultaneous determination of the recoil ion momenta.  Use and interpretation of these spectra are free from the complications of non-coincident total X-ray measurements that do not differentiate between the primary reaction channels.  The spectra offer the opportunity to test critically the ability of CX theories to describe such interactions at the  quantum orbital angular momentum level of the final projectile ion. To this end, new classical trajectory Monte Carlo calculations are compared here with the measurements. The current work demonstrates that modeling of cometary, heliospheric, planetary, and laboratory X-ray emission based on approximate state-selective CX models may result in erroneous conclusions and deductions of relevant parameters.

\end{abstract}





\section{Introduction}
X-ray and extreme ultraviolet (EUV) emission has been detected from more than 20 comets since the first observation by \citet{lis96}.  The charge exchange (CX) mechanism between highly charged solar wind (SW) minor heavy  ions and cometary neutrals suggested by \citet{cra97} is now recognized as the primary process responsible for the observed emission \citep [see, e.g.,][and references therein]{lis01,kra01,kra02,bei03,kha03,wil06,bod07,lis07}.  In the SWCX mechanism, electrons are captured from cometary neutrals by SW ions into excited states of the product ions, which may then decay radiatively and in the process emit X-ray radiation.  The SWCX mechanism has been invoked with various degrees  of sophistication to model and interpret cometary X-ray and EUV emission spectra 
\citep{hab97,weg98,sch00,kha03,otr07} and has been the subject of numerous reviews \citep{cra02,kra04,bha07,den08}.  It has been argued that cometary X-rays represent a potential tool to monitor not only cometary activity, but also the composition, velocity, and flux of the SW in regions that spacecraft cannot reach \citep{cra97,den97,sch00,bei01}.

It is now also recognized that heliospheric X-ray emission due to SWCX with H and He interstellar neutrals \citep [see, e.g.,][and references therein]{cox98,cra00,pep04,rob09},  and X-ray generation throughout the terrestrial magnetosheath  due to SWCX with geocoronal neutrals \citep [see, e.g.,][and references therein]{den97,cox98,cra09}  contribute to the soft X-ray background (SXRB). SWCX with H and O has also been proposed to account for the first definite detection of X-ray emission from the exosphere of Mars \citep{den06}. 


Understanding and accurately predicting these and related phenomena require novel experiments which simultaneously measure detailed (i.e., charge and quantum state-resolved) collision parameters in coincidence with consequent atomic energy de-excitation to elucidate the underlying chain of mechanisms in this CX induced X-ray emission, and ultimately, development of theoretical methods capable of broadly treating such interactions.  To this end, several 
experimental groups have carried out laboratory studies of relevant collision systems \citep [see, e.g.,][and references therein]{bei00,bei01,bei03,gre01,has01,gao04,ali05,bod06,maw07,all08,dju08}.  Of particular importance for accurate modeling is the ability to predict the $n\ell$-state-selective CX cross sections (i.e., to account for the distributions of the principal $n$ and angular momentum $\ell$ quantum numbers of the product projectile ions).  All previous modeling attempts to simulate cometary or  heliospheric X-ray spectra \citep{hab97,weg98,rig02,bei03,kha03,otr07,otr08} have adopted simple $n\ell$ empirical relations,  scalings from related collision systems, or fits to laboratory non-coincident total X-ray spectra.  It should be noted, however, that non-coincident laboratory spectra contain contributions from a variety of reaction channels such as single electron capture (SEC) and autoionizing and  non-autoionizing multiple-electron capture (MEC).  A superposition of several
reaction channels is also likely to occur in cometary, planetary, and heliospheric spectra. Therefore, a technique which is capable of differentiating between the primary reaction channels is required for the interpretation of such spectra.

In this letter, we report an experimental investigation of the $n$-state-selective hydrogen-like ion X-ray spectra following SEC in collisions of Ne$^{10+}$ with He, Ne, and Ar neutral targets at a laboratory frame collision velocity $v$ of 933 km s$^{-1}$ (4.54 keV nucleon$^{-1}$).  This velocity is at the upper end of the SW ion velocities.  The present interactions are close analogs of the interactions of heavy minor, multiply charged, SW ions with cometary, planetary, and heliospheric neutrals.  Specifically, the dominantly molecular constituents of cometary and planetary atmospheres (e.g., H$_2$O, CO$_2$) are simulated by gases of similar ionization potential and multielectron character (i.e., Ar, Ne).   Helium, being the second most abundant (15{\%}) interstellar neutral \citep{kou09}, is of direct relevance to heliospheric X-ray emission.  The present spectra are free from complications arising from the inability  of previously employed non-coincident total X-ray spectra to differentiate between the primary reaction channels.  Consequently they offer the opportunity to test critically the ability of theories to describe SWCX interactions at the $n\ell$ quantum level.

\section{Experiment}


Simultaneous cold-target recoil ion momentum spectroscopy (COLTRIMS) and X-ray spectroscopy were used for the triple-coincident detection of X-rays, scattered  projectiles, and target recoil ions. COLTRIMS has been reviewed by  \citet{dor00}, while the components of the experimental apparatus have been described elsewhere \citep{has99,ali05}.  Briefly, the $^{22}$Ne$^{10+}$ ions were provided by the University of Nevada,  Reno, 14 GHz ECR ion source, and guided to the collision chamber where they crossed supersonic target jets at $90^\circ$.  The target recoil ions resulting from the collisions were extracted by an electric field, at $90^\circ$ relative to the incident ions and jet,  and detected by a position-sensitive detector (PSD).  The scattered projectile ions were charge analyzed electrostatically and detected by another PSD where their impact positions provided their final charge states, while coincident time-of-flight (TOF) measurements between projectile and recoil ions provided the recoil ion charge states.  X-rays emitted at $90^\circ$ relative to the incident ions were detected by a windowless high-purity germanium detector, placed opposite the recoil detector.  Coincidences between projectile ions and X-rays ensured that all detected particles originated in the same collision event.  The full-width-at-half-maximum (FWHM) of the X-ray peaks is energy-dependent with a value of about 126~eV for the Ne$^{9+}$ Ly$\alpha$ line (1021.8 eV) and $\sim$133~eV for the Ly$\delta$ line (1307.7 eV).




\section{Results, theory, and discussion}

\begin{figure}
\includegraphics[scale=0.40]{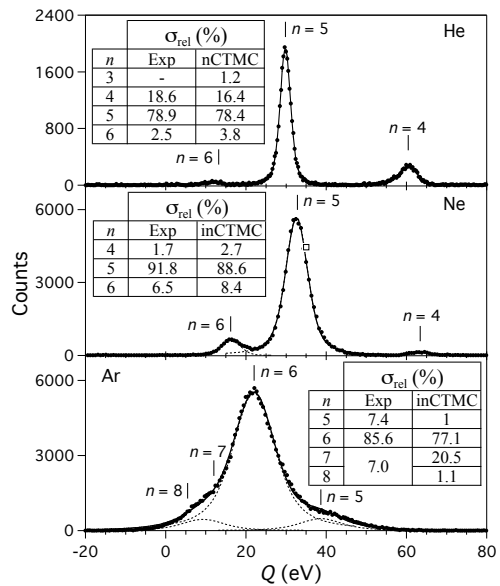}
\caption{Experimental SEC $Q$-value spectra for collisions of 933 km s$^{-1}$ $^{22}$Ne$^{10+}$ with He, Ne, and Ar targets.  Both experimental and theoretical $n$-state-selective relative cross sections $\sigma_n^{\rm rel}$ are tabulated for each target.}
\label{fig1}
\end{figure}

Since no electrons are directly ejected to the continuum for the considered collision energy, the change in electronic energy of the collision system, or the $Q$-value, is a direct measure of the projectile state population immediately following the collision.  $Q$-value 
spectra, therefore, provide the experimental $n$-state-selective  relative cross sections 
$\sigma_n^{\rm rel}$.  The $Q$-value for SEC is given by $Q\approx -(P_\parallel v+v^2/2)$ \citep{ali92}, where $P_\parallel$ is the longitudinal (i.e., parallel to the incident projectile direction) momentum transfer to the recoiling target. $P_\parallel$ of the recoil ions were determined from their TOF and impact positions on the PSD.  Figure \ref{fig1} displays the $Q$-value spectra for pure SEC  for the three considered collision systems with the $\sigma_n^{\rm rel}$ indicated.  These spectra were obtained using COLTRIMS-only measurements to acquire sufficient statistics for the accurate determination of $\sigma_n^{\rm rel}$.

Non-perturbative quantum mechanical treatments of highly charged ion collisions with multielectron targets are difficult. A recent study of Cl$^{7+}$ collisions with H \citep{zha07} demonstrates that such calculations are not routine, but require particular care, significant computational resources, and individual consideration. Therefore, we adopt a more tractable approach to model the measured interactions: the well-established classical trajectory Monte-Carlo (CTMC) method \citep{ols77}. The basic CTMC approach for bare ion collisions involving single-valence electron targets was discussed in \citet{has01}, while elaborations for multielectron targets in the context of total MEC was given in \citet{ali05}.  The He target was treated using the former method with the electron-nuclear charge interaction described by an effective charge $Z_{\rm eff}=1.6875$, designated as nCTMC (i.e., "CTMC" for "n" electrons bound by the atom's sequential ionization potentials).  For the Ne and Ar targets, six valence electrons within the independent particle model were considered  (inCTMC). The electron-atomic core interaction was treated with a  model potential. The standard microcanonical ensemble for the initial electron orbitals was filtered to remove all but the $p$ orbitals.

Examination of the experimental and CTMC $\sigma_n^{\rm rel}$ displayed in Fig. \ref{fig1} clearly shows that the agreement is excellent for both the He and Ne targets and reasonable for Ar.  Reasonable agreement between experiment and CTMC has been reported previously by \citet{cas96} for Ne$^{10+}$ and Ar$^{18+}$ on He at a slightly higher collision velocity ($\approx$ 1143 km s$^{-1}$).  That reasonable success has prompted the use of CTMC 
$\sigma_{n\ell}$ to simulate non-coincident total cometary and laboratory X-ray spectra \citep{otr07,otr08} under the assumption that the spectra are dominated by SEC. As we previously demonstrated \citep{ali05}, these spectra may contain significant contributions from MEC.  Furthermore, very high resolution (FWHM $\approx 10$ eV) total X-ray spectra using an X-ray microcalorimeter show clear signatures of MEC \citep{bei03}.  These and similar signatures in later work could not be accounted for by other SEC CTMC calculations \citep{otr07}, and the relative contributions to certain high energy emission lines from SEC and MEC remain in question \citep{war05,war08}.  Total X-ray spectra are, therefore, not the appropriate benchmarks for testing the validity of theoretical CX methods for simulating X-ray spectra.

\begin{figure}
\includegraphics[scale=0.40]{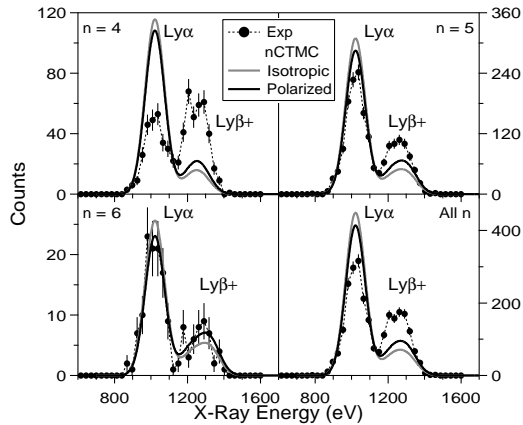}
\caption{Experimental and CTMC modeled $n$-state-selective and overall H-like Ne$^{9+}$ X-ray spectra following SEC by Ne$^{10+}$ from He. The modeled spectra assume isotropic emission (grey solid line) and maximum polarization for Ly$\beta +$ emission (black solid line).  The theoretical and experimental areas are normalized to each other to facilitate comparison.  The overall spectra preserve the experimental and theoretical 
$\sigma_n^{\rm rel}$.}
\label{fig2}
\end{figure}

\begin{figure}
\includegraphics[scale=0.40]{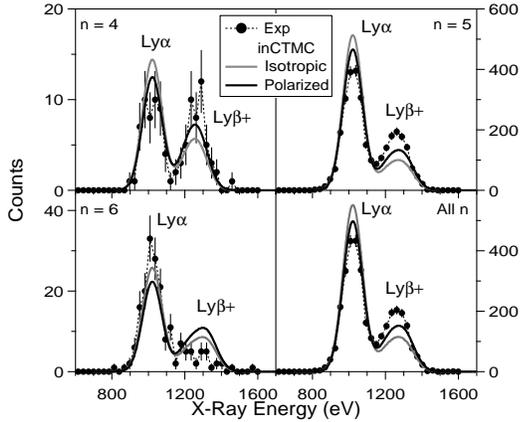}
\caption{Same as in Fig. \ref{fig2} but for SEC from Ne.}
\label{fig3}
\end{figure}

\begin{figure}
\includegraphics[scale=0.40]{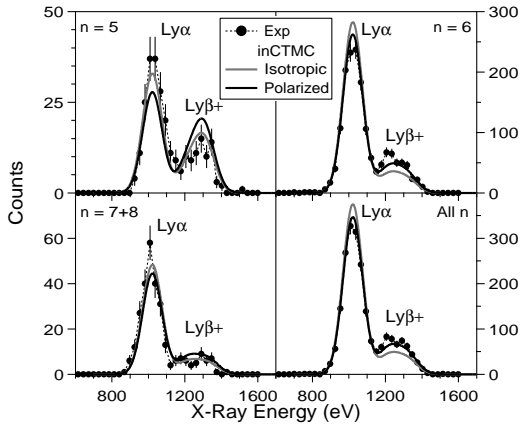}
\caption{Same as in Fig. \ref{fig2} but for SEC from Ar.}
\label{fig4}
\end{figure}

A major advantage of the simultaneous COLTRIMS and X-ray spectroscopic measurements is that it is not only possible to separate X-rays originating in pure SEC from those due to MEC 
\citep{ali05}, but it is also possible to obtain X-ray spectra corresponding to each populated $n$-level in the pure SEC channel as given in Figs. \ref{fig2}-\ref{fig4}.  The present measurements also provide the opportunity to test the validity of the theoretical methods, such as CTMC, for X-ray spectra modeling by assessing their ability to predict 
$\sigma_{n\ell}^{\rm rel}$ for SEC, and therefore the ability to simulate the most direct and simplest of cases (i.e., $n$-state-selective X-ray spectra).  To perform such a simulation, a radiative cascade model for the hydrogen-like Ne$^{9+}$ ion was constructed giving the relative yields of the different Lyman X-ray lines starting from a certain initial $n\ell$-state.  These  yields, together with the CTMC $\sigma_{n\ell}^{rel}$, the attenuation in the 300 {\AA} copper contact layer of the X-ray detector,  the variation of the FWHM of the Gaussian profiles with X-ray energy, and two X-ray radiation polarization scenarios, as explained below, have been taken into account in producing the CTMC $n$-state-selective X-ray spectra.

It is well known that X-ray emission following a CX collision is polarized \citep{ver85}.  Since the polarization rates are unknown in the present work, we considered two extreme scenarios.  The first assumes isotropic emission for all X-ray lines.  This assumption enhances the relative intensity of the Ly$\alpha$ to the $n\ge3 \rightarrow n=1$ (Ly$\beta +$) lines.  The other scenario  assumes isotropic emission for Ly$\alpha$ and adopts 100\% polarization for the Ly$\beta +$ emission that does not originate in or result from cascades passing through $s$-states.  The latter scenario maximizes the intensity of the Ly$\beta +$ lines relative to that of the  Ly$\alpha$ line.

The CTMC modeled $n$-state-selective and overall H-like neon X-ray spectra following SEC by Ne$^{10+}$ from the He, Ne, and Ar targets are shown in Figs.~\ref{fig2}, \ref{fig3}, and \ref{fig4}, respectively.  The theoretical and experimental line areas are mutually normalized for comparison while the overall spectra preserve $\sigma_n^{\rm rel}$.  The CTMC spectra reveal a general underestimation of the population of low-$\ell$ states, which give rise to Ly$\beta +$ lines, for the dominant $n$-level for each target.  This is also reflected in the overall SEC X-ray spectra. For minor capture channels, the behavior is less systematic, though agreement appears to be best with maximum polarization or for Ar targets.  The present results, based on the best available practical theoretical model, demonstrate that it remains a significant challenge to simulate accurately even the most simple cases.  Therefore, synthetic spectra of cometary, heliospheric, planetary, and laboratory total X-ray emission based on incompletely tested theoretical CX methods are not expected to be well founded.  This complements results from earlier investigations which showed that the dominance of SEC assumed in the models is also not justified \citep{ali05}.  

While the most recent X-ray emission simulations represent a major improvement over earlier efforts, which adopted equipartition or statistical angular momentum models \citep{hab97,weg98,sch00}, the present results suggest that agreement with observations is likely to be fortuitous.  Deductions of relevant parameters from such models should be considered as being associated with appreciable systematic uncertainties.

Considering the success of CTMC  in accounting for $\sigma_n^{\rm rel}$ as shown in Fig.~\ref{fig1}, it may seem surprising that it  underestimates the Ly$\beta +$ intensity for the
dominant $n$-levels.  CTMC, however, has a propensity towards a more statistical $\ell$-distribution (i.e., $\sigma_\ell \propto (2\ell + 1)$), at least for the lower $\ell$'s. Further, the agreement is seen to improve with decreasing first ionization potential: He (24.6 eV), Ne (21.6 eV), and Ar (15.8 eV).  In fact, the best agreement occurs for Ar which might be taken as a surrogate for the dominant species in cometary and planetary atmospheres responsible for the X-ray emission: H$_2$ (15.4 eV), CO (14.0 eV), CO$_2$ (13.7 eV), and H$_2$O (12.6 eV). The observation that the largest discrepancies occur for He suggests that the problem lies in electron-correlation effects which are known to be strongest for the He atom. Electron correlation is not sufficiently accounted for in the present CTMC calculations.

\section{Conclusions}

In summary, relative state-selective cross sections and X-ray spectra have been obtained from  triple-coincident measurements of X-rays, scattered projectiles, and target recoil ions which provide a test of CX theories at the quantum orbital angular momentum level. While improvements in X-ray simulations based on CX models have been made, the current results show that agreement for the relative state-selective cross sections does not necessarily imply agreement for the state selective X-ray spectra, and suggest that comparison of such models to observations of solar system X-rays may lead to faulty conclusions or parameter extractions.





\acknowledgments

Support is acknowledged from US DOE contract No. DE-AC52-07NA-27344 and NASA grant NNG06GB11G (PB), NASA grant NNH07AF12I (DRS), and NASA grants NNG05GD98G and NNX09AC46G (PCS).  RA acknowledges partial support from NASA EPSCoR, Nevada Astrophysics Program.  RA, PB, DRS, and PCS thank the NSF-funded Institute for Theoretical Atomic, Molecular, and Optical Physics at the Harvard-Smithsonian Center for Astrophysics for travel support.

\end{document}